\documentclass[prb,aps,preprint,superscriptaddress]{revtex4-1}
\usepackage[dvips]{graphicx}
\usepackage{rotating}
\begin{document}
\title{Evolution of electronic and magnetic properties in a series of iridate double perovskites Pr$_{2-x}$Sr$_x$MgIrO$_6$($x$ = 0, 0.5, 1.0)}
\author{Abhisek Bandyopadhyay}
\email[Corresponding author:] {msab3@iacs.res.in / abhisek.ban2011@gmail.com}
\affiliation{School of Materials Science, Indian Association for the Cultivation of Science, 2A \& 2B Raja S. C. Mullick Road, Jadavpur, Kolkata 700 032, India}
\author{Ilaria Carlomagno}
\affiliation{Dipartimento di Scienze, Universit\'{a} Roma Tre, Via della Vasca Navale, 84 I-00146 Roma, Italy}
\author{Laura Simonelli}
\affiliation{BL22 - CLAESS: Core Level Absorption \& Emission Spectroscopies Beamline responsible - Experiments Division, ALBA SYNCHROTRON LIGHT SOURCE, Ctra. BP 1413 km. 3,3 08290 Cerdanyola del Vall\`{e}s, Barcelona, Spain}
\author{M. Moretti Sala}
\affiliation{ESRF-The European Synchrotron, 71 Avenue des Martyrs, 38000 Grenoble, France}
\affiliation{Dipartimento di Fisica, Politecnico di Milano, P.zza Leonardo da Vinci 32, I-20133 Milano, Italy}
\author{A. Efimenko}
\affiliation{ESRF-The European Synchrotron, 71 Avenue des Martyrs, 38000 Grenoble, France}
\author{Carlo Meneghini}
\affiliation{Dipartimento di Scienze, Universit\'{a} Roma Tre, Via della Vasca Navale, 84 I-00146 Roma, Italy}
\author{Sugata Ray}
\affiliation{School of Materials Science, Indian Association for the Cultivation of Science, 2A \& 2B Raja S. C. Mullick Road, Jadavpur, Kolkata 700 032, India}


\date{\today}

\begin{abstract}
Spin-orbit coupling (SOC) plays a crucial role in magnetic and electronic properties of 5$d$ iridates. In this paper we have experimentally investigated the structural and physical properties of a series of Ir-based double perovskite compounds Pr$_{2-x}$Sr$_x$MgIrO$_6$($x$ = 0, 0.5, 1; abbreviated as PMIO, PSMIO1505 and PSMIO from now on). Interestingly, these compounds have recently been proposed to undergo a transition from the spin-orbit-coupled Mott insulating phase at $x$ = 0 to the elusive half-metallic antiferromagnetic (HMAFM) state with Sr-doping at $x$ = 1. However, our detailed magnetic and electrical measurements refute any kind of HMAFM possibility in either of the doped samples. In addition, we establish that within these Pr$_{2-x}$Sr$_x$MgIrO$_6$ double perovskites, changes in Ir-oxidation states (4+ for PMIO to 5+ for PSMIO via mixed 4+/5+ for PSMIO1505) lead to markedly different magnetic behaviors. While SOC on Ir is at the root of the observed insulating behaviors for all the three samples, the correlated magnetic properties of these three compounds develop entirely due to the contribution from local Ir-moments. Additionally, the magnetic Pr$^{3+}$ (4$f^2$) ions, instead of showing any kind of ordering, only contributes to the total paramagnetic moment. It is seen that the PrSrMgIrO$_6$ sample does not order down to 2 K despite antiferromagnetic interactions. But, the $d^5$ iridate Pr$_2$MgIrO$_6$ shows a sharp AFM transition at around 14 K, and in mixed valent Pr$_{1.5}$Sr$_{0.5}$MgIrO$_6$ sample the AFM transition is shifted to much lower temperature ($\sim$ 6 K) due to weakening of AFM exchange.
\end{abstract}

\maketitle

\newpage

\section{Introduction}
In contrast to the traditional wisdom of achieving uncorrelated wide band metals in 5$d$ iridates, spin-orbit coupling (SOC) plays a pivotal role in defining their complex magnetic and electronic ground states~\cite{sioprl}. Due to a delicate balance between SOC ($\lambda_0$), Coulomb correlation ($U$) and crystal field energy ($\Delta_{CFE}$), 5$d$ iridates particularly offer a promising avenue for hosting diverse physical properties~\cite{Review}. Strong SOC has been identified as the electronic reason of setting up an insulating band gap in Sr$_2$IrO$_4$ and other tetravalent iridates~\cite{sioprl,Review,6H5,6H6,6H7}. On the other hand, the relatively less explored pentavalent iridates (Ir$^{5+}$: 5$d^4$) stirred up a controversy about the origin of magnetism in them. Large SOC in low-spin Ir$^{5+}$ produces 15 possible organisations of spin-orbit coupled $J$ states (four electrons in three degenerate $t_{\text{2}g}$ orbitals each having two spin arrangements), with the atomic $J$ = 0 as lowest energy state~\cite{6H10,nagprl}, as shown in Fig. 1(a)-(c). But surprisingly, a pure nonmagnetic $J$ = 0 state has never been realized in any of the reported $d^4$ Ir-compounds till date~\cite{nagprl,6H11,6H13,nagprbdp}. Actually the strength of SOC and Hund's exchange together determine the relative stability of the $LS$/$jj$ coupled multiplet states (Fig. 1(b)). One plausible route for magnetic moment generation in these $d^4$ iridates has been assigned to Van-Vleck-type intrasite singlet-triplet excitations ($J$ = 0 $\rightarrow$ $J$ = 1) due to comparable energy scales between superexchange (mediated by complex Ir-O--O-Ir paths) and SOC-driven singlet-triplet gap~\cite{6H15}. Otherwise, another most prominent factors against the observation of a nonmagnetic state  could be the solid state effects, such as large bandwidth of the 5$d$ orbitals, ligand-Ir charge transfer, non-cubic crystal field and intersite Ir-Ir hopping which always act against the atomic SOC effect~\cite{6H11,6H15,nagprbdp6,nagprbdp8}, and hence, produce finite magnetic moments~\cite{nagprl,nagprbdp,6H24}. Infact, there is an active debate running currently regarding the trueness of the proposal of excitonic magnetism in these cases against the ground state magnetism, originated by hopping and other solid state effects~\cite{nagrixsprl}
\par
In this backdrop, a recent theoretical claim of half-metallic antiferromagnetism (HMAFM) appeared for a $d$$^4$ iridate double perovskite (DP) compound PrSrMgIrO$_6$~\cite{prbhmafm}. The half-metallicity (HM) has been proposed by first assuming the dominance of exchange splitting which prevents the mixing of spin-up and spin-down bands and ensures 100\% spin polarization at Fermi energy $E_F$. Hence the strength of the SOC has been considered to be comparatively negligible and Ir energy levels have been treated within $LS$ coupling limit. On the other hand, the vanishing net macroscopic magnetic moment has been described by AFM coupling of Pr$^{3+}$ (4$f^2$: 2 $\mu_B$) with the Ir$^{5+}$ (5$d^4$) moment in the $LS$ coupling limit. On the other hand, the undoped double perovskite Pr$_2$MgIrO$_6$ is predicted to be a ferrimagnetic Mott insulator~\cite{prbhmafm,exppmio}. Clearly, a few controversies prevailed with this description, such as, (i) despite having strong SOC on Ir$^{5+}$, prediction of half-metallicity in the two Sr-doped compounds and development of large magnetic moment/Ir$^{5+}$ ion in the PrSrMgIrO$_6$ compound, and (ii) unlike the other existing perovskite/double perovskite compounds, non-Kramer Pr$^{3+}$ ion at the $A$-site of the present set of double perovskites is predicted to be exchange coupled to the magnetic $B$-site to provide the magnetic ground state. Thus, in order to sort out the aforementioned contradictions, we have synthesized Ir-based DP compounds Pr$_{2-x}$Sr$_x$MgIrO$_6$ ($x$ = 0, 0.5, 1; identified as PMIO, PSMIO1505 and PSMIO from now onwards) to verify the magnetc and electronic properties of them. Interestingly, subtle differences in IrO$_6$ octahedral distortions, and also changes in the Ir-valence state upon Sr-doping show profound influence on their physical properties.
\par
Here in this paper we show that, neither half-metallicity nor any kind of long-range AFM/ferrimagnetic ordering is observed in either of the doped compounds. Finally, our experimental observations reveal the actual scenario: Pr$^{3+}$ does not undergo any kind of magnetic ordering or spin freezing down to the lowest measuring temperature in all these compounds, instead, the Pr$^{3+}$ only contributes to the total paramagnetic moment. Like in other Ir-based oxides ~\cite{nagprbdp,6H11,lzioprb,lszioprb}, the ground state magnetic properties of these compounds are solely influenced by the spin-orbit coupled $J$-states of Ir. In PrSrMgIrO$_6$ (PSMIO), featureless magnetic susceptibility, AFM interactions, and no sign of magnetic ordering down to 2 K are evident. Further, presence of small magnetic moment at the Ir-site drives this system away from ideal $J$ = 0 limit. On the other hand, the undoped Pr$_2$MgIrO$_6$ (PMIO) (Ir$^{4+}$: 5$d^5$, magnetic species) undergoes a long-range AFM transition at around 14 K. In mixed valent Pr$_{1.5}$Sr$_{0.5}$MgIrO$_6$ (PSMIO1505) compound, the AFM transition gets weakened with the introduction of Ir$^{5+}$. On top of such magnetizations, all the three compounds exhibit SOC-driven insulating ground states.


\section{Experimental Section}
Polycrystalline Pr$_{2-x}$Sr$_x$MgIrO$_6$($x$ = 0, 0.5, 1) samples have been synthesized by conventional solid state reaction technique. Stoichiometric amounts of high purity ($>$ 99.9\%) Pr$_2$O$_3$, SrCO$_3$, MgO, and IrO$_2$ powders were thoroughly mixed in an agate mortar. This mixture has been calcined initially at 850$^{\circ}$ C for 12 hours in air to decompose carbonates and finally sintered at 1250$^{\circ}$ C for 48 hours in air with few intermediate grindings. The structural characterization of all the samples was performed using a Bruker AXS: D8 Advance x-ray diffractometer. The X-ray-diffraction (XRD) data were analyzed by using the Rietveld technique and refinements were done by FULLPROF program~\cite{fullprof}. To verify homogeneity and any off-stoichiometry in the sample, Energy Dispersive X-Ray (EDX) Analysis was also performed using Field Emission Scanning Electron Microscope (FE-SEM, JEOL, JSM-7500F). Electrical resistivity was measured by standard four-probe method within a temperature range of 200 - 400 K in a lab-based resistivity set up. Magnetization measurements in the temperature range 2 - 300 K and in magnetic fields up to $\pm$5 T were performed in a superconducting quantum interference device (SQUID) magnetometer (Quantum Design). Ir $L_3$-edge X-ray absorption fine structure (XAFS) experiments of the respective samples have been performed at the BL22-CLAESS beamline of ALBA (Barcelona, Spain)~\cite{CLAESS} synchrotron radiation facility at room temperature in standard transmission geometry. Data treatment and quantitative analysis of EXAFS (extended x-ray absorption fine structure) were carried out using the freely available Demeter package~\cite{artemis,ravel} (Athena \& Arthemis programs) using atomic clusters from crystallographic structure to i) individuate the single and multiple scattering contributions relevant for the quantitative EXAFS data refinement and ii) calculate (FEFF6L program)  theoretical amplitude and phase functions required to calculate the theoretical EXAFS  curve assuming Gaussian disorder. The X-ray photoemission spectroscopy (XPS) measurements were carried out using OMICRON electron spectrometer, equipped with SCIENTA OMICRON SPHERA analyzer and Al $K_{\alpha}$ monochromatic source with an energy resolution of 0.5 eV. Before collecting the spectra the sample surface was sputtered with argon ion bombardment for each of these samples to remove any kind of surface oxidization effect and the presence of environmental carbons in the pelletized samples. The collected spectra were processed and analyzed with Kolxpd program. Further, the RIXS (Resonant Inelastic X-ray Scattering) measurements at the Ir $L_3$-edge of PrSrMgIrO$_6$ sample were performed at the ID20 beamline of European Synchrotron Radiation Facility (ESRF) using $\pi$-polarized photons and a scattering geometry with 2$\theta$ $\simeq$ 90$^{\circ}$ to suppress elastic scattering. A spherical, diced Si(844) analyzer was used in a Rowland circle of 2 m radius in combination with a custom-built hybrid pixel detector, having an overall energy resolution of $\approx$ 29 meV at the Ir $L_3$ edge in this configuration. Apart, Ir $L_3$-RIXS of Pr$_2$MgIrO$_6$ and Pr$_{1.5}$Sr$_{0.5}$MgIrO$_6$ samples was measured at the CLEAR spectrometer of BL22-CL{\AE}SS beamline of ALBA (Barcelona, Spain) synchrotron radiation facility with an energy resolution $\approx$ 1 eV.

\section{Results and Discussion}
\subsection{Structure from X-ray diffraction}
Rietveld refined powder x-ray diffraction (XRD) patterns, obtained from polycrystalline samples of Pr$_{2-x}$Sr$_x$MgIrO$_6$ ($x$ = 0, 0.5, 1.0) at room temperature, confirm pure single phase with monoclinic $P$2$_{1/n}$ space group for all three samples, as indicated in Fig. 2(a). Further, the Energy Dispersive X-ray (EDX) analysis ensures that these three samples are chemically homogeneous and cation stoichiometry is retained at the target composition, {\it i.e.,} Pr:Sr:Mg:Ir being very close to 1:1:1:1 and 1.5:0.5:1:1 ratios for PSMIO and PSMIO1505 respectively, while the undoped PMIO attains nearly 2:1:1 (Pr:Mg:Ir) ratio within the given accuracy of the measurement. Presence of superlattice reflection at around 2$\theta$ = 20$^{\circ}$ of the XRD patterns suggest significant Mg/Ir ordering at the $B$-site of these three DP compounds. Although the XRD refinements clearly infer full Mg/Ir chemical order at the $B$-site (see the respective occupancies as indicated in Table-I) of two Sr-doped compounds, $\sim$ 3-4\% Mg/Ir disorder remains evident from the XRD refinement of undoped Pr$_2$MgIrO$_6$ case (see Table-I). Lattice parameters, atomic positions, site-occupancy, along with the goodness factors for all the three samples are listed in Table-I. Due to large size mismatch between Sr$^{2+}$ (1.44{\AA}) and Pr$^{3+}$ (1.12{\AA}), Pr/Sr layered ordering is expected at the $A$-site~\cite{woodward} of PSMIO sample. As a result, the O-Ir-O bond angles (within a single IrO$_6$ octahedral unit), sitting closer to Pr(Sr)-layer, would be reduced (increased) with respect to the ideal 90$^{\circ}$ for perfect cubic symmetry. This brings higher degree of rotational distortion in the IrO$_6$ octahedra of PSMIO compared to the other two compounds, as displayed in Fig. 2(d)-(f) and also tabulated in Table-2. So, the effect of noncubic crystal field, arising from IrO$_6$ octahedral rotation, would be larger in case of PSMIO. In addition to this, each Ir ion is acted upon by local noncubic crystal field in all these compounds due to the presence of three oxygen sites, and consequently, three different Ir-O bond lengths [Fig. 2(d)-(f)]. The structural distortions often bring complex magnetism in iridates~\cite{6H11,6Hprb}. Also, the frustrated equilateral triangular network, usually formed out of $B$-site cations in cubic double perovskites, is replaced by isosceles triangles in all the three samples [see Fig. 2(g)-(i)]. This shall cause different extent of geometric frustration in these compounds.

\subsection{Local structure from EXAFS}
The EXAFS data analysis provide finest details about the local coordination geometry and local chemical order (the  antisite defects) which are complementary to structural information obtained from XRD analysis (Rietveld refinement) which only probes long range coherent structural features. The local antisite disorder in the double perovskite structure often largely influences the magnetic response~\cite{ownprb,6H42,JAP_calafeiro6}. Therefore, to confirm the local coordination around (IrO$_6$) and also the chemical order (antisite defects) at both the $A$- and $B$-sites, the Ir $L_3$-edge EXAFS data (see Fig. 3) has been analyzed in the $R$ space in the 1-6 {\AA} region for the two Sr-doped samples, while 1-4 {\AA} region for the undoped one (see Fig. 3(a)-(f)). Unlike the two Sr-doped compounds, the weak nature of the EXAFS signal of Pr$_2$MgIrO$_6$, specially in the 5-6 {\AA} region (fourth shell) of the Fourier transform data (Fig. 3(d)), restricts us to carry out satisfactory fitting of the FT pattern above third shell ($>$ 4 {\AA}) for this sample. Such a weak signal must be attributed to the larger octahedral tilting distortions of Pr$_2$MgIrO$_6$ in contrast to the two doped samples, likely in agreement with the XRD results (see Fig. 2(b) and Table-2). We applied a multi-shell data refinement procedure~\cite{tanushree_jmcc,payel_prb} in order to access next neighbour structural information, relevant to describe the chemical order and antisite defects. The obtained results are summarized in Table-3. The EXAFS data analysis confirms almost negligible Mg/Ir chemical disorder ($\approx$ 3\%, 1\% and 0.6\% for PMIO, PSMIO1505 and PSMIO respectively) for all the samples with comparable local interatomic distances with the XRD refinements. These disorder percentages obtained from EXAFS analysis are very much consistent with the XRD refinements [discussed in Sec. III-(A) and shown in Table-1]. EXAFS is more suitable for probing the true nature of local chemical order and is not so suitable for probing bulk order, still the values obtained from EXAFS analysis are mentioned for the sake of completeness. Further, our analysis suggests that every Ir ion appears to find 4 Pr and 4 Sr as nearest neighbour cations in PSMIO, confirming homogeneous Pr/Sr distribution at the $A$-site. On the other hand, each Ir sees 8 Pr for Pr$_2$MgIrO$_6$, while 6 Pr/2 Sr as nearest neighbour cations around Ir for Pr$_{1.5}$Sr$_{0.5}$MgIrO$_6$, as expected for the desired compositions in the respective cases. It should be noted that in the fitting approximation, we did not consider the multiple scattering (MS) contributions for the undoped Pr$_2$MgIrO$_6$ sample (see Fig. 3(a) and Table-3), as the largely distorted structure and also little higher Mg/Ir disorder reduce the focussing effect, thereby, weakening the MS terms and thus, addition of MS paths did not improve the fitting in this case.

\subsection{Ir-valance state from XANES and core level X-ray Photoelectron Spectroscopy (XPS)}
The stoichiometric formulaes of Pr$_{2-x}$Sr$_x$MgIrO$_6$ ($x$ = 0, 0.5, 1.0) samples suggest that Ir should be in the 4+ and 5+ oxidation states in case of PMIO and PSMIO respectively, while PSMIO1505 should carry 4.5+ valence (mixture of 4+ and 5+) of Ir, in order to maintain the charge balance. The Ir-oxidation state has been of central importance in the magnetism of Ir-based compounds, as Ir$^{4+}$/Ir$^{6+}$ ions are magnetic~\cite{sioprl,lzioprb,sriro3prb,EJIC_Ir6+}, while Ir$^{5+}$ should ideally be nonmagnetic ($J$ = 0) in the $jj$ coupling scenario. So to confirm the charge states, Ir $L_3$-edge XANES spectra for the three samples have been collected and shown in Fig. 4(a) (i)-(iii) along with the respective theoretical fittings by fixing the background at the $arctangent$ shape and the peak width at 2.5 eV for all the three samples. These spectra clearly exhibit a systematic chemical shift [as indicated by orange dotted line in Fig. 4(a)] as well as appearance of rich asymmetric curve shape with Sr-doping, indicating gradual increase of Ir-oxidation state in these compounds~\cite{irxanes1,irxanes2}. The corresponding second derivative curves, representative of the {\it white line} (2$p$ $\rightarrow$ 5$d$ transition) feature, are presented in Fig. 4(b).
Well resolved doublet features in the white line spectra of all these compounds indicate the 2$p$ $\rightarrow$ $t$$_{2g}$ (low energy feature) and 2$p$ $\rightarrow$ $e$$_g$ (higher energy peak) transitions. The peak shape as well as a gradual development of the peak feature corresponding to 2$p$ $\rightarrow$ $t$$_{2g}$ transition, supported further by the enhancement of the area under the solid green curve [shown by the respective XANES spectra fitting in Fig. 4(a)(i)-(iii)], confirms the expected Ir-oxidation states~\cite{irxanes1,irxanes2} (4+ for $x$ = 0 to 5+ for $x$ = 1.0 via an intermediate between 4+ and 5+ for $x$ = 0.5). On the contrary, the peak shape and the peak intensities [area under each solid blue curve corresponding to the three samples, highlighted in Fig. 4(a)] of the 2$p$ $\rightarrow$ empty $e$$_g$ transition remain nearly unchanged (the minor changes in the peak area are within the error bar of the experiment) irrespective of the change in Ir-oxidation state in these three compounds. In addition, the shape of the features corresponding to 2$p$ $\rightarrow$ $t$$_{2g}$ transitions for the three samples [see Fig. 4(b)] matches very well with the observation of previously reported Ir-based double perovskites~\cite{irxanes1}.

In addition, the Ir 4$f$ core level XPS spectra were collected and fitted using a single spin-orbit split doublet for Pr$_2$MgIrO$_6$ and PrSrMgIrO$_6$ compounds, while two spin-orbit split doublets were required for the fitting of the 25\% Sr-doped compound [see Figs. 4(c)-(e)]. The energy positions of the respective 4$f_{7/2}$ and 4$f_{5/2}$ features in the doublets along with their spin-orbit separations (around 3.05-3.1 eV) for the three samples, confirm pure 5+ and pure 4+ charge states of Ir in PSMIO and PMIO compounds respectively, while mixed 4+/5+ valance states for the 25\% Sr-doped sample~\cite{nagprl,nagprbdp,majumdar227}.

\subsection{Resonant Inelastic X-ray Scattering (RIXS)}
Representative low resolution Ir $L_3$-edge RIXS spectra for the three samples (data of both PMIO and PSMIO1505 have been collected from CL{\AE}SS beamline of ALBA where PSMIO was measured at ESRF) have been plotted after (0, 1) normalization as a function of energy loss at $T$ = 300 K, shown in Fig. 5(a) in the same panel for the sake of comparison. The largest energy loss features ($\sim$ 6 eV and $\sim$ 9 eV) correspond to charge transfer excitations from the O 2$p$ bands to unoccupied Ir $t_{\text{2}g}$ and empty $e_g$ bands respectively~\cite{nagprbdp33}. The feature observed at $\sim$ 3.5-3.6 eV represents electron excitation from $t_{\text{2}g}$ to $e_g$ orbital, indicating the crystal field energies of these samples. The slightly reduced value of $t_{\text{2}g}$ $\rightarrow$ $e_g$ crystal field excitation in PSMIO sample compared to the other two compounds [see Fig. 5(a)] is due to further splitting of the crystal field-driven Ir $t_{\text{2}g}$ and $e_g$ orbitals caused by the IrO$_6$ octahedral distortion [see Section-III (A)]. On the other hand, the feature corresponding to O 2$p$ to unoccupied Ir $t_{\text{2}g}$ transition gets consistently intensified in the doped samples [see Fig. 5(a)-(c)], as Sr-doping introduces Ir$^{5+}$ (5$d^4$) ions which creates more number of holes in the $t_{\text{2}g}$ orbital, thereby, enhancing the transition probability and consequently resulting in a sharp feature at $\sim$ 6 eV for the pentavalent iridate DP PSMIO, shown in Fig. 5(a). Although the rising feature (at $\sim$ 6 eV) is consistent with increasing number of Ir $t_{\text{2}g}$ holes in these compounds upon Sr-doping, the discrepancy in the order of their intensities is possibly due to different experimental setups in ESRF (ID23 beamline for PSMIO) and ALBA (CL{\AE}SS beamline for the other two samples) synchrotron facilities. While ESRF setup pushes energy resolution at the expenses of flux, the opposite is applicable to the ALBA's setup. In addition, within similar measurement configuration the increased intensity of the O 2$p$ to Ir $e_g$ transition (at $\sim$ 9 eV) in Pr$_{1.5}$Sr$_{0.5}$MgIrO$_6$ compared to the undoped Pr$_2$MgIrO$_6$ [shown in Fig. 5(b) and (c)] further supports higher degree of IrO$_6$ octahedral distortions in PSMIO1505 relative to the PMIO case [discussed in Section-III(A) and shown in Fig. 2(d) and (e)] and also points to the difference in local environments around the Ir-O octahedra due to the existence of mixed Ir$^{4+}$/Ir$^{5+}$ valence states in PSMIO1505, contrary to the pure 4+ charge state of PMIO. Consequently, the different extent of transition probabilities between the differently splitted Ir energy levels of these two compounds causes intensity variation in the absorption spectra.

\subsection{Electrical Resistivity and XPS valance band spectra}
The temperature variation of electrical resistivity ($\rho$($T$)) for the three samples are shown in Fig. 6. Upon cooling, resistivity increases continuously for all the samples, indicating insulating behaviors of them. Further, the $\rho$($T$) curves could be modeled by Mott variable range hopping (VRH) mechanism in three dimensions~\cite{vrh} as, $\rho$($T$) $\sim$ $\exp$($T$$_0$/$T$)$^{1/4}$, shown in the insets to Figs. 6(a)-(c). The valance band XPS spectra for these three samples were further collected and the results are summarized in Figs. 6(d)-(f). As displayed, complete absence of density of states at the Fermi level affirm the charge-gapped electronic ground states for all the three compounds. Thus our observation of insulating nature in all the three samples immediately refutes the claim for half-metallicity in the Sr-doped compounds~\cite{prbhmafm}, suggesting the dominance of SOC over the exchange splitting, similar to the other reported 5$d$ iridate double perovskites~\cite{nagprbdp,lszioprb}.

\subsection{Magnetization}
Next we have investigated the nature of magnetization of these systems. The dc magnetic susceptibility $\chi$($T$) of the undoped PMIO sample [Fig. 7(a)], measured at 100 Oe applied magnetic field, shows sharp antiferromagnetic (AFM) transition near around 14 K. A Curie-Weiss (C-W) fit [using equation $\chi$ = $\chi$$_0$ + $C$/($T$ - $\Theta$$_{CW}$); $\chi$$_0$ is the temperature independent paramagnetic susceptibility while $C$ and $\Theta$$_{CW}$ represent Curie constant and Curie-Weiss temperature respectively] to the field-cooled susceptibility data [shown by blue solid line in Fig. 7(a)], in the temperature range 100 K $\leq$ $T$ $\leq$ 300 K, provides an effective paramagnetic (PM) moment $\mu$$_{eff}$ $\sim$ 5.2 $\mu$$_B$/f.u and a negative $\Theta$$_{CW}$ $\sim$ -36.3 K, suggesting AFM interactions within the compound. In order to understand the contribution of magnetic Pr$^{3+}$ ion on magnetism and to estimate the spin-orbit coupled Ir$^{4+}$ moment, we have considered the temperature dependent {\it dc} magnetization of an isostructural Ir-double perovskite La$_2$MgIrO$_6$ (LMIO)~\cite{lzioprb} and we also synthesized the same LMIO compound for better comparison of Curie-Weiss analysis between PMIO and LMIO in the same temperature range and applied magnetic field. Due to similar lattice constants of LMIO with PMIO, any change of the Ir-Ir interactions due to lattice change could be presumed to be negligibly small. Further, same oxidation states of both Pr and La ensures that for both the compounds the Ir-oxidation state will remain same. The AFM transition temperature ($\sim$ 12.5 K) of LMIO~\cite{lzioprb} resembles the AFM transition of the PMIO. Further, the effective paramagnetic moment, obtained from the Curie-Weiss fit on field-cooled $\chi$($T$) data of LMIO in the 100-300 K temperature range (not shown in the figure), takes a value of $\sim$ 1.36 $\mu$$_B$/Ir$^{4+}$. Using this moment value for Ir$^{4+}$ and considering the paramagnetic moment for a Pr$^{3+}$ ion in the $LS$ coupling limit to be 3.58 $\mu$$_B$~\cite{exppmio,ln2naoso6}, the theoretically calculated effective magnetic moment for PMIO becomes,
\begin{equation}
\mu_{eff} = \sqrt{2\times(\mu_{eff})_{Pr^{3+}}^2 + (\mu_{eff})_{Ir^{4+}}^2} \mu_B/f.u
\end{equation}
\begin{equation}
\Rightarrow \mu_{eff} = \sqrt{2\times(3.58)^2 + (1.36)^2} \mu_B/f.u = 5.24 \mu_B/f.u,
\end{equation}
This value is in extremely good agreement with our C-W fit. Actually, Pr$^{3+}$ is a non-Kramer ion, and therefore, sufficiently low symmetry crystal field at the $A$-site of the perovskite/double perovskite structure completely removes the degeneracy of the $J$ ground multiplet of Pr$^{3+}$ and results in nine singlets~\cite{prvo3,prvo312,prmno3prb2004,prmno3prb2006}. As a result of which, Pr$^{3+}$ residing at the $A$-site of these perovskite/double perovskite compounds~\cite{ln2naoso6,prvo3,prvo312,prmno3prb2004,ln2nairo6} does not possess any kind of magnetic coupling with the magnetic $B$-site. So it should be quite convincing to claim that the ordered magnetic behaviors of all the three samples in the present study will solely be influenced by the spin-orbit coupled Ir-moments, while Pr$^{3+}$ should only act as the paramagnetic background over the entire measuring temperature range.
\par
The dc magnetic susceptibility for the 25\% Sr-doped sample (PSMIO1505) in 1000 Oe applied field is presented in Fig. 7(b). The susceptibility curves remain nearly featureless without any ZFC/FC divergence, similar to the observation of most $d$$^4$ iridates~\cite{nagprbdp,6H13,syioprb,slioperov}. Only a very weak AFM-like kink appears at $\sim$ 6 K [inset to Fig. 7(b)]. This suggests weakening of magnetic interactions. Curie-Weiss fitting on the 1000 Oe field-cooled $\chi$($T$) data in the temperature range 100-300 K provides an effective paramagnetic moment, $\mu$$_{eff}$ $\sim$ 4.5 $\mu$$_B$/f.u and a negative $\Theta$$_{CW}$ of $\sim$ -38.5 K. Considering a nonmagnetic ground state for Ir$^{5+}$, the theoretically calculated effective magnetic moment for PSMIO1505 would be,
\begin{equation}
\mu_{eff} = \sqrt{1.5\times(\mu_{eff})_{Pr^{3+}}^2 + 0.5\times(\mu_{eff})_{Ir^{4+}}^2 + 0.5\times(\mu_{eff})_{Ir^{5+}}^2}
\end{equation}
\begin{equation}
\Rightarrow \mu_{eff} \approx \sqrt{1.5\times(3.58)^2 + 0.5\times(1.36)^2} \mu_B/f.u.
\end{equation}
\begin{equation}
\Rightarrow \mu_{eff} = 4.49 \mu_B/f.u.
\end{equation}
This value again agrees very well with our C-W fitting. Actually, Sr$^{2+}$-doping introduces Ir$^{5+}$ (5$d^4$), which increases the spatial separation between the magnetic Ir$^{4+}$ ions due to increased density of the Ir$^{5+}$ ions upon hole-doping. Thus, the strength of magnetic exchange interaction between the magnetic Ir$^{4+}$ ions is suppressed with respect to the undoped compound, resulting in a weakening of the AFM transition in PSMIO1505. Also the greater extent of exchange frustration within the isosceles Ir triangular network [see Fig. 2(h)] of this compound compared to the undoped one [see Fig. 2(g)] possibly dilutes the effect of AFM transition in the present case. Like in PMIO, on top of the Ir-magnetism, the Pr$^{3+}$ sublattice only enhances the total PM moment of this system.
\par
Finally, the temperature variation of the 5000 Oe {\it dc} magnetic susceptibility ($\chi$($T$)) curves for the PSMIO sample is presented in Fig. 7(c). Absence of any feature confirms no magnetic long-/short-range ordering down to 2 K, like in other $d^4$ Ir-compounds~\cite{6H13,syioprb}. The C-W fit on the field-cooled data, in the temperature range 150-300 K, gives a $\Theta$$_{CW}$ of $\sim$ -38.6 K. The effective magnetic moment ($\mu$$_{eff}$ $\sim$ 3.89 $\mu$$_B$/f.u), obtained from this fit, is slightly higher than 3.58 $\mu$$_B$/Pr$^{3+}$. So the remaining excess moment ($\sim$ 0.3 $\mu$$_B$) is getting developed obviously at the Ir$^{5+}$-site, driving the system away from the expected $J$ = 0 nonmagnetic ground state. The presence of few percent of magnetic Ir$^{4+}$/Ir$^{6+}$ ions, as the possible origin of moment development~\cite{6H13,lsmgiro6_zaac}, could be refuted in the present case from the Ir $L_3$-edge XANES analysis (discussed in the XANES portion) and also the Ir 4$f$ core level XPS data (discussed in XPS portion). Further, negligible Mg/Ir chemical disorder ($<$ 1\% as discussed in the EXAFS section) at the $B$-site of this DP strongly discards any chance of moment generation due to enhanced Ir-Ir exchanges because of Mg/Ir antisite defects. So one might consider the effect of non-cubic crystal field (see structural discussion), which reduces the effect of atomic SOC, as the origin of weak Ir-moments~\cite{6H11} in this compound by redistributing the spin-orbit coupled $J$ multiplets. In addition, the another highly decisive factor for the development of such small finite moment on individual Ir$^{5+}$ could be due to intersite real Ir-Ir hopping causing delocalization of the intrasite Ir$^{5+}$ holes and thus, deviating from a perfect atomic $d^4$ configuration~\cite{nagprbdp,nagrixsprl}, causing magnetic ground state. According to A. Nag {\it et al.}~\cite{nagrixsprl}, it has been argued that even moderate hopping, present in the systems like cubic Ba$_2$YIrO$_6$~\cite{nagprbdp}, can be suspected as the origin of atomic SOC rescaling and subsequent development of finite magnetic moment. The deviation from C-W law below 150 K suggests development of short-range correlations between the Ir-moments~\cite{irti2prb}. Despite having significant AFM interactions (negative $\theta_{CW}$ value), this sample does not possess ordering down to 2 K at least possibly due to geometric frustration arising from the isosceles Ir-triangles of three nearly identical Ir-Ir bond distances [see Fig. 2(i)]. Thus, the Ir-Ir AFM exchange interactions are expected to be of nearly similar strength for nearest neighbours on all of the Ir-sites in PSMIO, preventing this compound from magnetic order. The field-dependent magnetization $M$($H$) curves for PSMIO (not shown) show neither hysteresis nor any saturation in any of the temperatures. As evident in the inset to Fig. 7(c), the 2 K Arrot plot ($M^2$ versus $H$/$M$) renders intercept on the negative $M^2$ axis, clearly discarding the presence of spontaneous magnetization {\it viz-a-viz} FM components~\cite{y2ir2o7_jmmm} in this PSMIO sample.
\subsection{High resolution RIXS of PSMIO}
Although a nonmagnetic $J$ = 0 ground state is ideally expected at the Ir$^{5+}$-site of PSMIO compound from single atomic perspective, presence of a finite magnetic moment on individual Ir$^{5+}$ ion has been confirmed from magnetization measurements. So, it is important to comment on the trueness of atomic $J$ state description in this double perovskite. Consequently, the high-resolution Ir $L_3$-edge RIXS spectra (measured at $T$ = 20 K and 300 K) of PrSrMgIrO$_6$ sample have been collected and illustrated in Fig. 8 (a). During experiment, the incident photon energy
was kept fixed at 11.216 keV, which was found to enhance the low energy inelastic features of the $J$ multiplet excitations. In order to get deeper insight into these features, the high resolution low energy RIXS spectra of perfectly $B$-site ordered cubic double perovskite Ba$_2$YIrO$_6$ were measured within same technical specifications~\cite{nagprbdp} as PSMIO at $T$ = 20 and 300 K, and the subsequent results are represented in Fig. 8(a) along with PSMIO.  Like in Ba$_2$YIrO$_6$ (BYIO)~\cite{nagprbdp}, we observe three similar inelastic peaks below 1.5 eV here in PSMIO. Although the shape and energy positions of these three peaks appear similar in both the samples, subtle changes in these inelastic RIXS features are clearly evident, as demonstrated by intensity enhancement and shift in energy position of the first feature (indicated by greenish ellipses of Fig. 8(a)) as well as development of prominent shoulder in the higher energy side of the second peak (shown by red shaded arrows in Fig. 8(a)) in the PSMIO sample contrary to Ba$_2$YIrO$_6$ case. Clearly, both these compounds belong to double perovskite crystal structure with rock-salt ordered Y-Ir / Mg-Ir arrangements at the $B$-site while the only difference lies in the space group symmetry of the respective crystal structures. It is known that a perfectly cubic $Fm\bar{3}m$ is adopted by BYIO while much lower monoclinic crystal symmetry becomes applicable in PSMIO, and as a result, hopping pathways (see Fig. 8(b) and (c)) for PSMIO suffer significant octahedral tilting distortion in terms of Ir-O-Mg bond angles in contrast to 180$^{\circ}$ Ir-O-Y connectivity for BYIO. On top of it, monoclinic symmetry driven local noncubic crystal field around IrO$_6$ octahedra of PSMIO removes the Ir $t_{2g}$ degeneracy and consequently rearranges the Ir energy levels, opposite to the ideal cubic crystal field for BYIO~\cite{nagprbdp}. In such a scenario, we may qualitatively infer that the aforementioned differences in RIXS features for PrSrMgIrO$_6$ relative to the Ba$_2$YIrO$_6$ should be due to the dissimilar Ir-Ir hopping connectivities (Fig. 8(b) and (c)) and also the influence of noncubic crystal distortions in PSMIO. Indeed, both hopping and noncubic crystal field have strong impact on the effective SOC strength in $d^4$ Ir-systems~\cite{nagrixsprl} and therefore, defining the spin-orbit coupled Ir energy levels from the perspective of atomic $J$ picture only, as has been the widely accepted scenario till very recently~\cite{byioprbrixs}, becomes insufficient, as revealed by A. Nag {\it et al.}~\cite{nagrixsprl} and A. Revelli {\it et al.}~\cite{sciencerixs} recently. So it is very clear that precise estimation of SOC strength on Ir within atomic limit is not at all a reasonable approach because the low energy Ir $L_3$ inelastic RIXS features would be the outcome of intersite hopping, local noncubic crystal distortions and several other electronic factors. So to elucidate the effective strength of SOC and the resulting new $J$ states (Fig. 8 (a)) in PSMIO, further Full Multiplet calculations will be required which should include all possible electronic and solid state effects.
\section{Conclusions}
In conclusion, we have performed a systematic study of the structural and physical properties of Pr$_2$MgIrO$_6$ and its hole-doped counterparts Pr$_{2-x}$Sr$_x$MgIrO$_6$ ($x$ = 0.5, 1.0) via X-ray diffraction, X-ray absorption fine structure, High and low resolution RIXS, dc magnetization and electrical resistivity measurements. We find insulating charge sector of these three compounds confirming a leading role of SOC on the Ir-site. However, the effect of atomic SOC is reduced due to the presence of significant noncubic crystal distortion and the ground state magnetism gets affected by hopping among Ir-Ir sites, producing small correlated moments on every Ir-site of the $d$$^4$ iridate double perovskite PrSrMgIrO$_6$ and hence, causing a breakdown of the ideal atomic $J$ = 0 picture. No sign of magnetic ordering was found down to 1.85 K for this $d$$^4$ iridate indicating at frustration parameter ($f$ = $\frac{\Theta_{CW}}{T_N}$ with $T_N$ is the lowest measuring temperature here) being $>$ 20, while exchange interactions between the strongly magnetic Ir$^{4+}$ ions of the undoped Pr$_2$MgIrO$_6$ compound result in long-range AFM transition at low temperatures. Compared to the undoped one, Pr$_{1.5}$Sr$_{0.5}$MgIrO$_6$, with Ir-valence in between 4+ and 5+, exhibits weakening of the magnetic exchange interaction due to half substitution of the magnetic Ir$^{4+}$ ions by nonmagnetic Ir$^{5+}$ ions. As a result, the AFM transition is suppressed. On top of this, Pr$^{3+}$ does not take part in correlated magnetism of either of these compounds, instead, it only acts as a paramagnetic background to enhance the total paramagnetic moment of all these samples.
\section{Acknowledgements}
AB thanks CSIR, India for fellowship. SR thanks Department of Science and Technology (DST) [Project No. WTI/2K15/74] for support, and Jawaharlal Nehru Centre for Advanced Scientific Research from DST-Synchrotron-Neutron project, for performing experiments at ESRF (Proposal No. HC-2872). AB and SR thank CL{\AE}SS beamline of ALBA (Barcelona, Spain) synchrotron radiation facility. Authors also thank TRC-DST of IACS for providing experimental facilities.

\newpage

\begin{table}
\caption{All the samples are refined within a single crystallographic phase. Monoclinic $P$2$_{1/n}$ space group is taken for all the compositions.\\
(a)Pr$_2$MgIrO$_6$($300$ K): $a$ = 5.504(8) {\AA}, $b$ = 5.659(8) {\AA}, $c$ = 7.835(4) {\AA}; $\alpha$ = $\gamma$ = 90$^{\circ}$, $\beta$ = 89.9901$^{\circ}$\\
$R$$_{p}$ = 20.0, $R$$_{wp}$= 19.1, $R$$_{exp}$= 10.26, and $\chi$$^{2}$=3.48\\
(b)Pr$_{1.5}$Sr$_{0.5}$MgIrO$_6$($300$ K): $a$ = 5.538(7) {\AA}, $b$ = 5.610(7) {\AA}, $c$ = 7.851(9) {\AA}; $\alpha$ = $\gamma$ = 90$^{\circ}$, $\beta$ = 90.0055$^{\circ}$\\
$R$$_{p}$ = 15.6, $R$$_{wp}$= 14.4, $R$$_{exp}$= 7.08, and $\chi$$^{2}$=4.11\\
(c)PrSrMgIrO$_6$(300 K): $a$ = 5.565(7) {\AA}, $b$ = 5.574(2) {\AA}, $c$ = 7.865(8) {\AA}; $\alpha$ = $\gamma$ = 90$^{\circ}$, $\beta$ = 90.0212$^{\circ}$\\
$R$$_{p}$ = 21.0, $R$$_{wp}$ = 19.4, $R$$_{exp}$ = 10.56, and $\chi$$^{2}$ = 3.37\\}
\resizebox{9cm}{!}{
\begin{tabular}{| c | c | c | c | c | c | c |}
\hline Sample & Atoms & occupancy &  $x$ & $y$ & $z$ & $B$ ($\times$ 10$^{3}${\AA}$^2$) \\\hline
  & Pr & 1.0 & 0.4928(6) & 0.0516(9) & 0.2497(6) & 2.3(7)\\
  & Mg1 & 0.964 & 0 & 0 & 0 & 0.9(5)\\
  & Ir1 & 0.036 & 0 & 0 & 0 & 0.9(5)\\
  & Ir2 & 0.964 & 0.5 & 0.5 & 0 & 1.2(1)\\
 Pr$_2$MgIrO$_6$ & Mg2 & 0.036 & 0.5 & 0.5 & 0 & 1.2(1)\\
  & O1 & 1.0 & 0.1966(2) & 0.2655(5) & 0.0497(3) & 4.7(6)\\
  & O2 & 1.0 & 0.6144(8) & 0.4756(4) & 0.2561(5) & 4.7(6)\\
  & O3 & 1.0 & 0.2717(2) & 0.7937(7) & 0.0552(8) & 4.7(6)\\\hline
  & Pr & 0.75 & 0.4961(7) & 0.0434(1) & 0.2506(5) & 1.8(5)\\
  & Sr & 0.25 & 0.4961(7) & 0.0434(1) & 0.2506(5) & 1.8(5)\\
  & Mg & 1.0 & 0 & 0 & 0 & 0.8(2)\\
Pr$_{1.5}$Sr$_{0.5}$MgIrO$_6$ & Ir & 1.0 & 0.5 & 0.5 & 0 & 1.5(3)\\
  & O1 & 1.0 & 0.2015(2) & 0.2479(8) & 0.0221(2) & 7.2(8)\\
  & O2 & 1.0 & 0.5873(0) & 0.4669(5) & 0.2673(1) & 7.2(8)\\
  & O3 & 1.0 & 0.2788(5) & 0.7952(6) & 0.0640(9) & 7.2(8)\\\hline
  & Pr & 0.5 & 0.4978(1) & 0.0259(9) & 0.2504(0) & 2.9(4)\\
  & Sr & 0.5 & 0.4978(1) & 0.0259(9) & 0.2504(0) & 2.9(4)\\
  & Mg & 1.0 & 0 & 0 & 0 & 1.2(6)\\
PrSrMgIrO$_6$ & Ir & 1.0 & 0.5 & 0.5 & 0 & 2.3(1)\\
  & O1 & 1.0 & 0.2053(8) & 0.2602(2) & -0.008 & 6.6(4)\\
  & O2 & 1.0 & 0.5728(0) & 0.4745(8) & 0.2557(8) & 6.6(4)\\
  & O3 & 1.0 & 0.3041(8) & 0.8097(2) & 0.0463(5) & 6.6(4)\\
\hline
\end{tabular}
}
\end{table}

\begin{table}
\caption{Estimation of rotational and tilting distortions of the IrO$_6$ octahedral unit in the form of deviated bond angles for all three samples.}
\resizebox{9cm}{!}{
\begin{tabular}{| c | c | c | c | c |}\hline
  & Connectivities & Pr$_2$MgIrO$_6$ & Pr$_{1.5}$Sr$_{0.5}$MgIrO$_6$ & PrSrMgIrO$_6$ \\\hline
  & O1-Ir-O2 & 88.897$^{\circ}$ & 87.907$^{\circ}$ & 82.067$^{\circ}$ \\
Rotational distortion & O1-Ir-O3 & 90.756$^{\circ}$ & 87.152$^{\circ}$ & 82.892$^{\circ}$ \\
  & O2-Ir-O3 & 91.944$^{\circ}$ & 88.038$^{\circ}$ & 89.355$^{\circ}$ \\\hline
  & Ir-O3-Mg & 151.271$^{\circ}$ & 147.071$^{\circ}$ & 147.282$^{\circ}$ \\
Tilting distortion & Ir-O2-Mg & 143.526$^{\circ}$ & 150.315$^{\circ}$ & 155.361$^{\circ}$ \\
  & Ir-O1-Mg & 152.801$^{\circ}$ & 165.262$^{\circ}$ & 166.902$^{\circ}$ \\\hline
\end{tabular}
}
\end{table}

\begin{table}
\caption{Local structure parameters as obtained from the EXAFS analysis of Ir $L_3$-edge for the three samples. In order to reduce correlation among the parameters, constraints among the parameters were applied, namely, $x$ as the fraction of IrPr pairs, {\it i.e.,} N$_{IrPr}$ = 8$x$ and N$_{IrSr}$ = 8(1 $-$ $x$) for the doped samples, and $xx$ as the fraction of IrOMg configurations, {\it i.e.,} N$_{IrOMg}$ = 6$\ast$$xx$ and N$_{IrOIr}$ = 6(1 $-$ $xx$) for all the samples. The fixed or constrained values are labeled by `$\ast$'. The absolute mismatch between the experimental data and the best fit are $R$$^{2}$ = 0.022, 0.025 and 0.011 for PSMIO, PSMIO1505, and PMIO respectively.}
\resizebox{9cm}{!}{
\begin{tabular}{| c | c | c | c | c |}
\hline Sample & Shell & $N$ & $\sigma$$^{2}$ ($\times$ 10$^{2}${\AA}$^2$) & $R$({\AA}) \\\hline
   & Ir-O & 6.0$\ast$ & 0.21(4) & 2.01(3) \\
   & Ir-Pr1 & 2.0$\ast$ & 0.27(3) & 3.24(5) \\
   & Ir-Pr3 & 4.0$\ast$ & 0.27(3)$\ast$ & 3.39(3) \\
  Pr$_2$MgIrO$_6$ & Ir-Pr4 & 2.0$\ast$ & 0.27(3)$\ast$ & 3.51(5) \\
   & Ir-Mg & 5.8 & 0.68(1) & 3.86(9) \\
   & Ir-Ir (antisite defect) & 0.2 & 0.68(1)$\ast$ & 3.86(9) \\\hline
   & Ir-O & 6.0$\ast$ & 0.22(8) & 1.98(6) \\
   & Ir-Pr1 & 2.0$\ast$ & 0.79(2)$\ast$ & 3.22(5)\\
   & Ir-Sr3 & 1.96 & 0.79(2) & 3.41(2) \\
  Pr$_{1.5}$Sr$_{0.5}$MgIrO$_6$ & Ir-Pr3 & 4.04 & 0.79(2)$\ast$ & 3.41(2) \\
   & Ir-Mg (SS) & 5.92 & 0.74(5) & 3.93(8) \\
   & Ir-O-Mg (MS-3 legs) & 11.78 & 0.74(5)$\ast$ & 4.00(2) \\
   & Ir-Ir (4$^{th}$ shell) & 12.0$\ast$ & 0.47(5) & 5.56(6) \\\hline
   & Ir-O & 6.0$\ast$ & 0.34(6) & 1.95(9) \\
   & Ir-Sr & 3.91 & 0.78(5) & 3.34(9) \\
  PrSrMgIrO$_6$ & Ir-Pr & 4.09 & 0.78(5) & 3.32(8) \\
   & Ir-Mg (SS) & 5.96 & 0.56(1) & 3.91(4) \\
   & Ir-O-Mg (MS-3 legs) & 11.9 & 0.56(1)$\ast$ & 3.95(6) \\
   & Ir-Ir (4$^{th}$ shell) & 12.0$\ast$ & 0.57(5) & 5.55(6) \\
\hline
\end{tabular}
}
\end{table}

\newpage

\begin{figure}
\caption{(color online) Redistribution of $d^4$ orbitals of an Ir$^{5+}$ ion under octahedral crystal field (a), then atomic SOC (b) to form spin-orbit coupled multiplet states, and finally a nonmagnetic $J$ = 0 state under strong SOC limit (c) in the single particle picture.}
\end{figure}

\begin{figure*}
\caption{(color online) (a) Rietveld refined XRD patterns of all the synthesized samples. Open black circles represent the experimental data and continuous red line represents the calculated pattern. The blue line represents the difference between the observed and calculated pattern while the vertical green lines indicate the Bragg position for all the samples. The refined crystal structures for (b) Pr$_2$MgIrO$_6$ and (c) PrSrMgIrO$_6$. The rotational distortions (change in O-Ir-O bond angles) within IrO$_6$ octahedral unit for Pr$_2$MgIrO$_6$ (d), Pr$_{1.5}$Sr$_{0.5}$MgIrO$_6$ (e), and PrSrMgIrO$_6$ (f) samples. In addition, extent of geometric frustration caused by Ir-triangles are shown for Pr$_2$MgIrO$_6$ (g), Pr$_{1.5}$Sr$_{0.5}$MgIrO$_6$ (h), and PrSrMgIrO$_6$ (i) compounds.}
\end{figure*}

\begin{figure}
\caption{(colour online) Ir $L$$_3$-edge $k$$^2$ weighted experimental EXAFS data (shaded black circles) and the corresponding best fits (red solid line) for Pr$_2$MgIrO$_6$ (a) in the $k$ range: 3-14 {\AA}, Pr$_{1.5}$Sr$_{0.5}$MgIrO$_6$ (b) in the $k$ range: 3-16 {\AA} and PrSrMgIrO$_6$ (c) in the $k$ range: 3-16 {\AA}. The contributions from the individual single  and multiple scattering paths (solid colored line) and the residual [$k$$^{2}$$\chi$$_{exp}$-$k$$^{2}$$\chi$$_{th}$] (open cyan dots) are also shown for these three samples, vertically shifted for clarity. The Fourier Transforms of the respective experimental data (shaded black circles) and the theoretical (solid red line) curves for Pr$_2$MgIrO$_6$ (d), Pr$_{1.5}$Sr$_{0.5}$MgIrO$_6$ (e) and PrSrMgIrO$_6$ (f) samples; the magnitude ($|$$FT$$|$) and the imaginary parts ($Imm$) are also indicated; vertically shifted for clarity.}
\end{figure}

\begin{figure}
\caption{(colour online) (a) Ir $L_3$-edge XANES (X-ray Absorption Near Edge Spectroscopy) spectra (shaded black circles) for Pr$_2$MgIrO$_6$ (a-i), Pr$_{1.5}$Sr$_{0.5}$MgIrO$_6$ (a-ii), and PrSrMgIrO$_6$ (a-iii), along with their respective fittings (coloured solid line). (b) Second derivative curves of the respective normalized absorption spectra, indicating white line feature. Further, Ir 4$f$ core level XPS spectra (shaded black circles) along with the fitting (red solid line) for Pr$_2$MgIrO$_6$ (c), Pr$_{1.5}$Sr$_{0.5}$MgIrO$_6$ (d), and PrSrMgIrO$_6$ (e) samples.}
\end{figure}

\begin{figure}
\caption{(colour online) (a)-(c) Low-resolution high-energy RIXS features for the three samples.}
\end{figure}

\begin{figure}
\caption{(colour online) (a)-(c) Temperature dependent electrical resistivity variations for the three samples. Inset: corresponding Mott VRH fitting; Further, XPS valance band spectra for (d) Pr$_2$MgIrO$_6$, (e) Pr$_{1.5}$Sr$_{0.5}$MgIrO$_6$, and (f) PrSrMgIrO$_6$ samples.}
\end{figure}

\begin{figure}
\caption{(colour online) Zero Field cooled (open circles) and Field cooled (shaded circles) {\it dc} magnetic susceptibilities as a function of temperature ($\chi$($T$) under 100 Oe applied field for Pr$_2$MgIrO$_6$ (PMIO) (a); (b) The temperature dependent {\it dc} magnetic susceptibility in both ZFC (open green circles) and FC (shaded green circles) modes at $H$ = 1000 Oe field for Pr$_{1.5}$Sr$_{0.5}$MgIrO$_6$ (PSMIO1505); Inset: expanded view of the 50 Oe $\chi$($T$) curve for the same sample. (c) ZFC (open pink circles) and FC (shaded pink circles) {\it dc} magnetization curves for the PrSrMgIrO$_6$ (PSMIO) sample at 5000 Oe applied magnetic field; Inset: The Arrot plots ($M^2$ versus $H$/$M$ curve) at $T$ = 2 K.}
\end{figure}

\begin{figure}
\caption{(colour online) (a) High-resolution RIXS spectra at $T$ = 20 K (upper panel) and 300 K (lower panel) for the PrSrMgIrO$_6$ sample, clearly showing the low energy inelastic features; Also, the 20 K and 300 K low energy RIXS features for another double perovskite Ba$_2$YIrO$_6$ are further plotted in the respective figures for comparison. Ir-Ir hopping pathways for PrSrMgIrO$_6$ (b) and Ba$_2$YIrO$_6$ (c) samples, mediated via corner shared (Mg/Y)O$_6$ octahedral units.}
\end{figure}

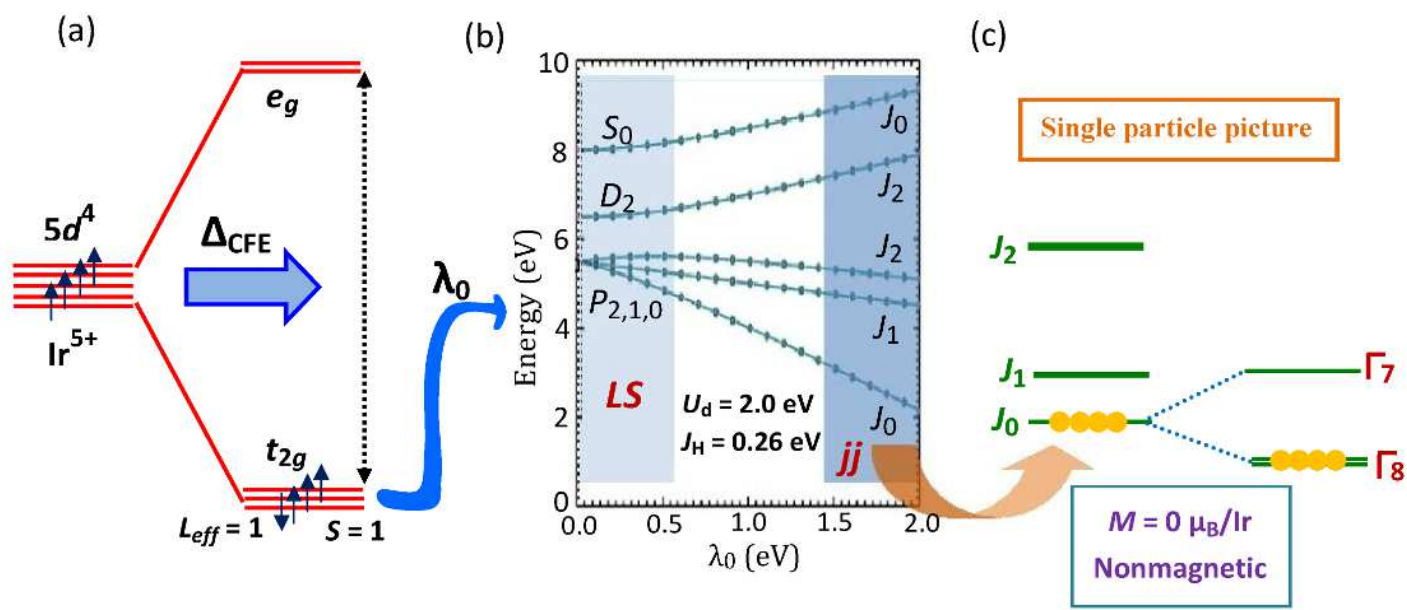
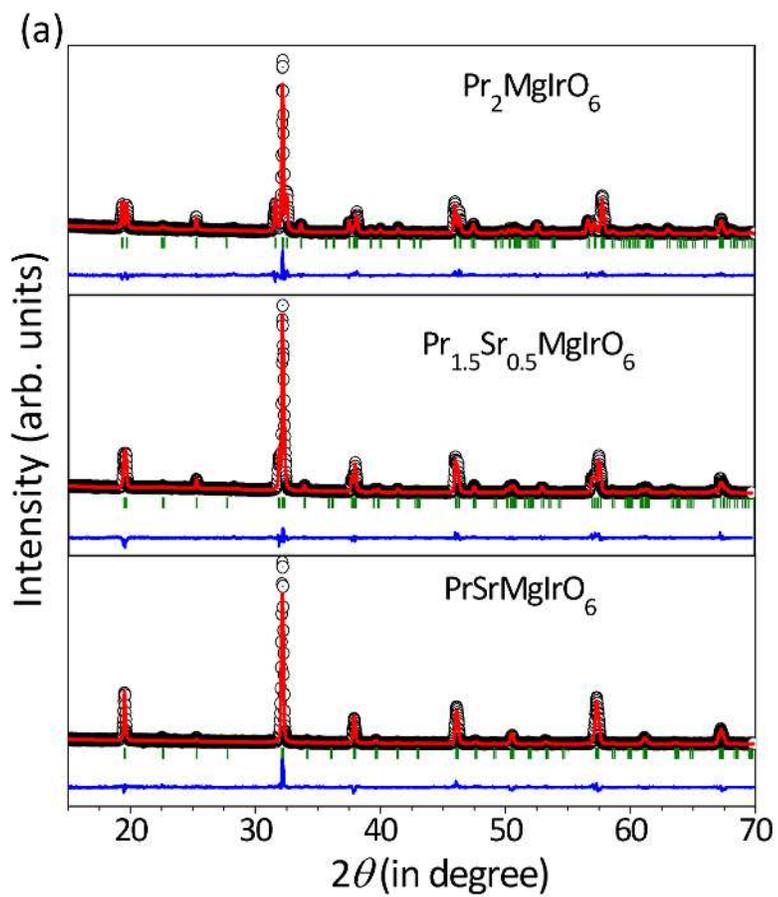
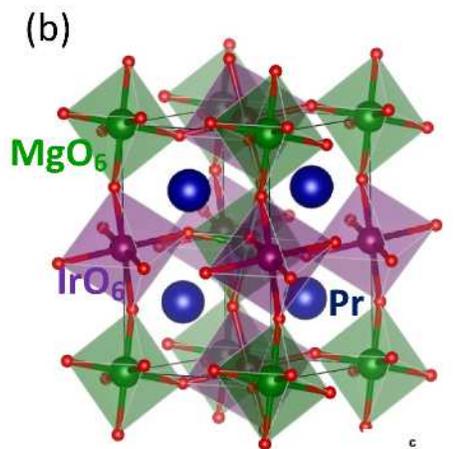
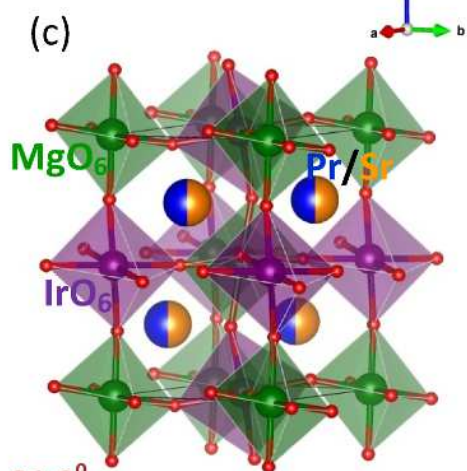
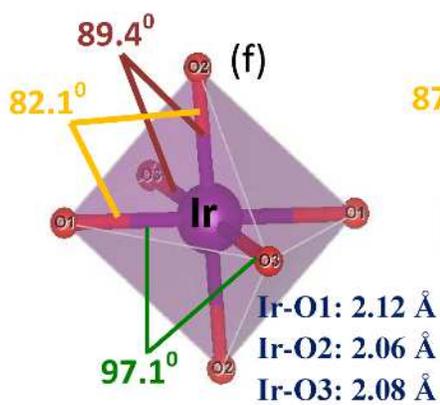
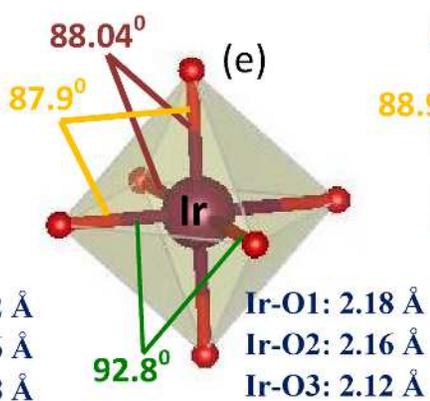
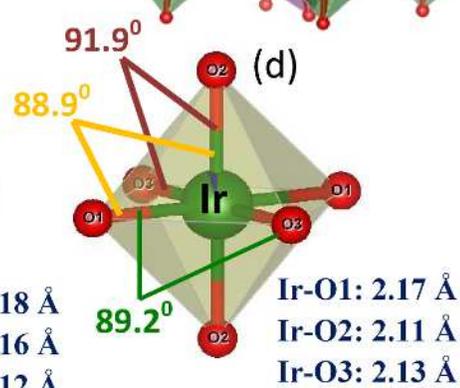
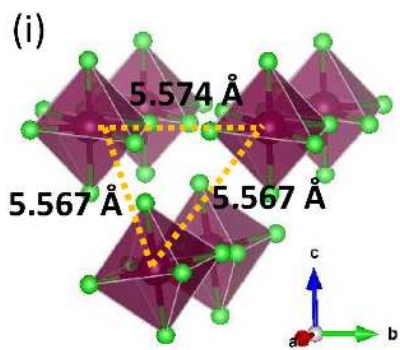
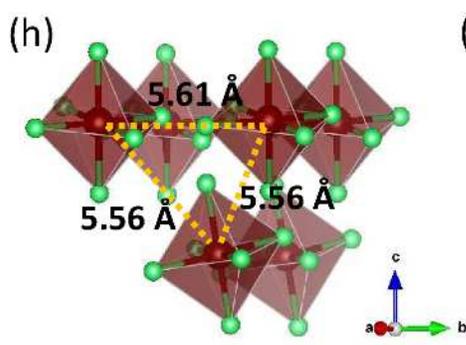
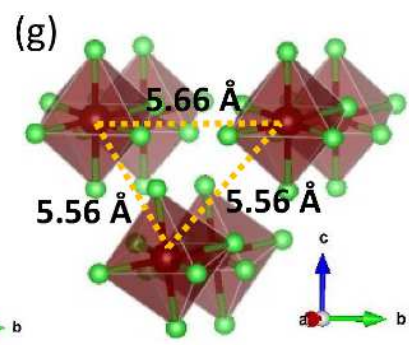
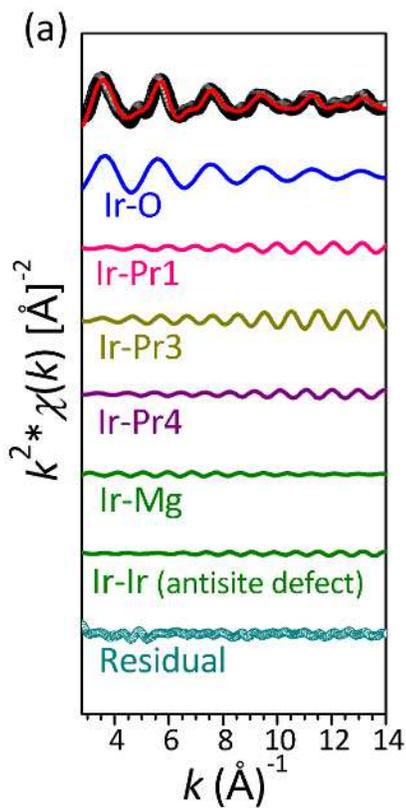
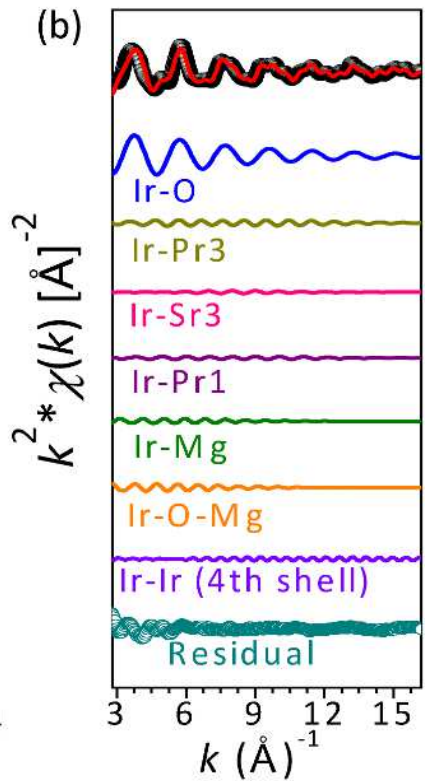
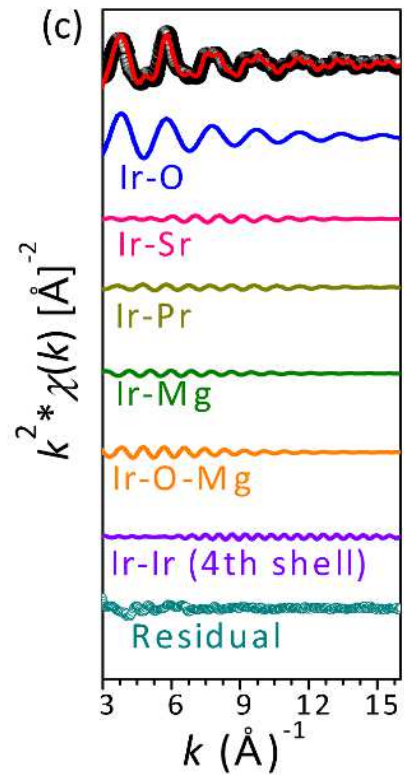
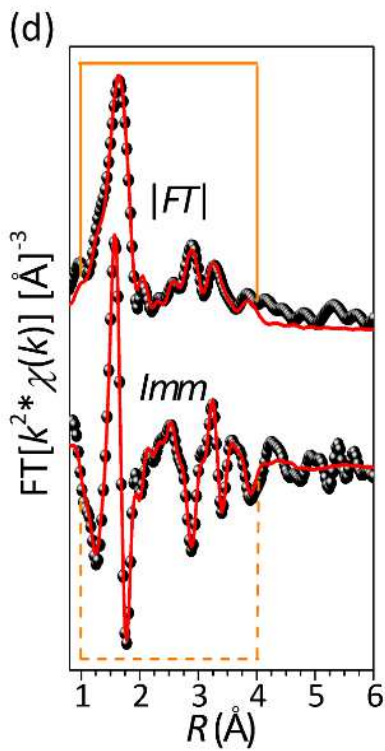
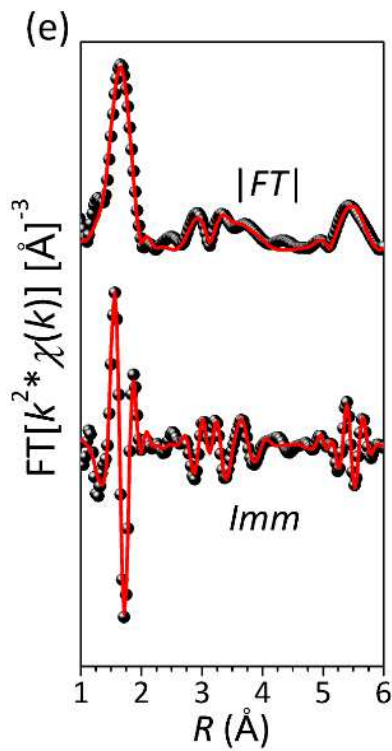
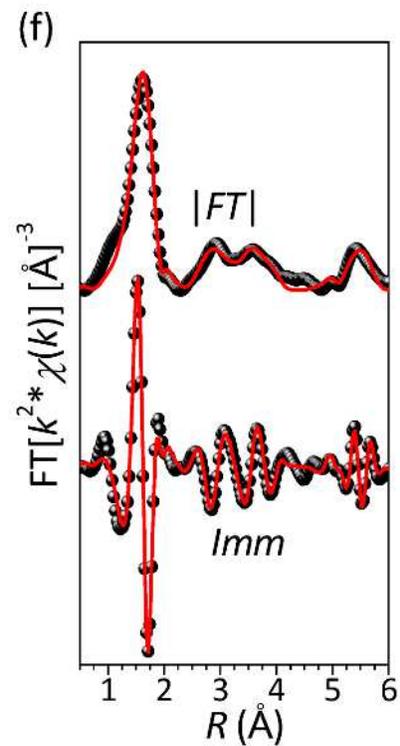
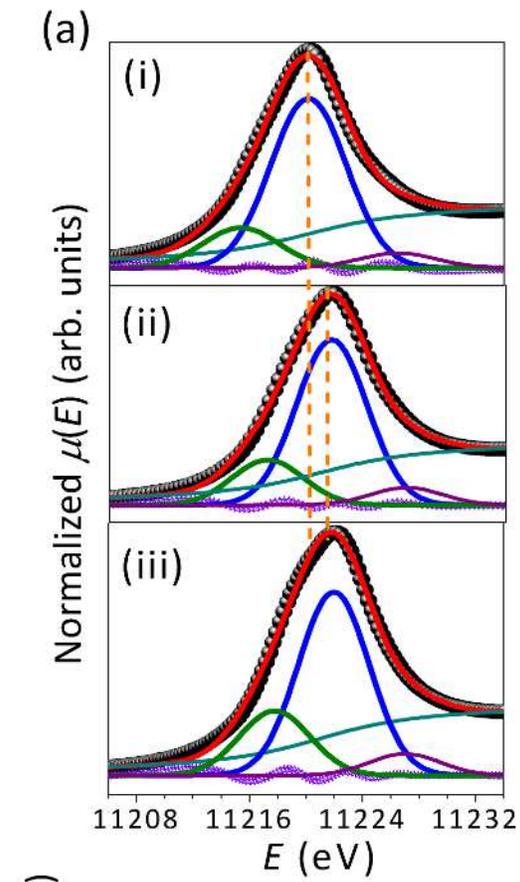
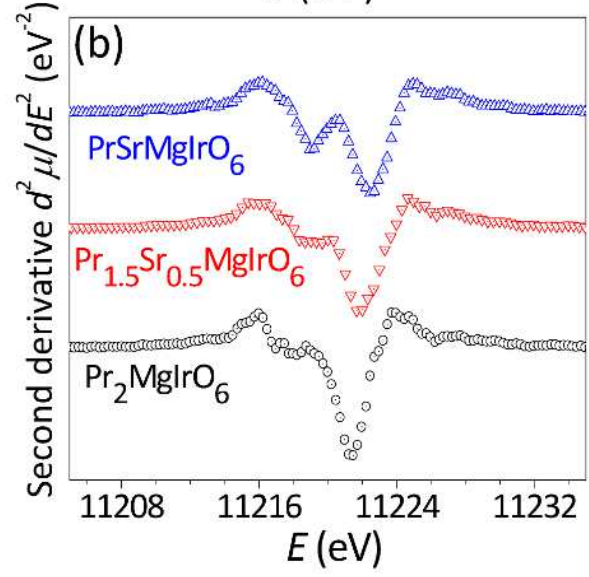
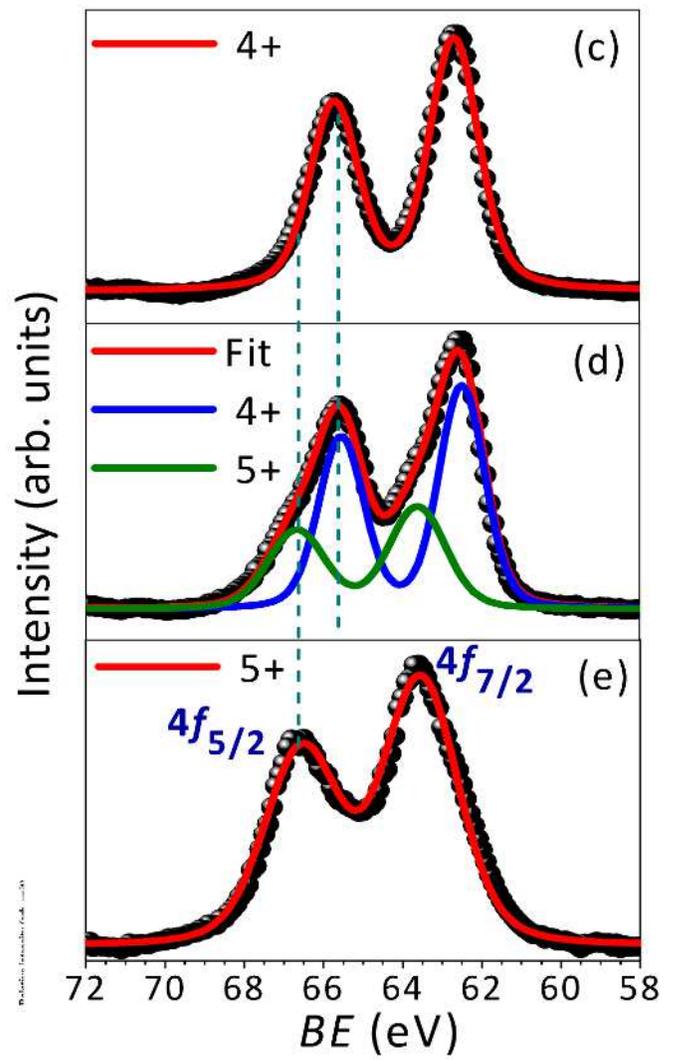
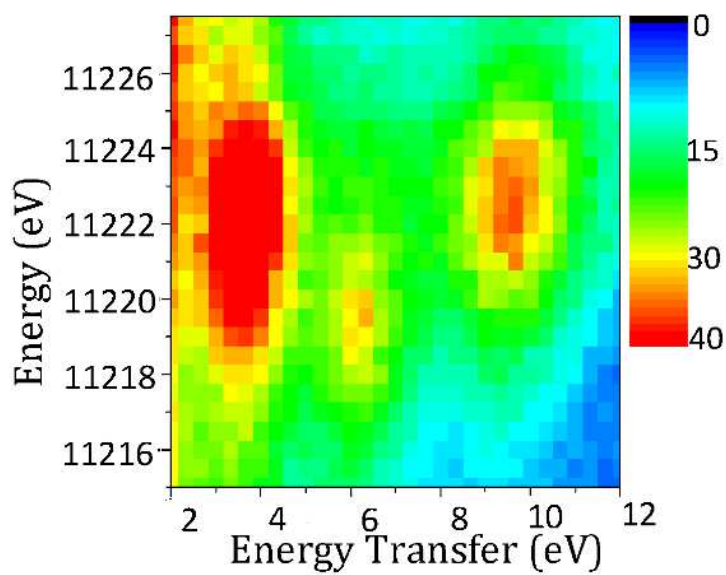
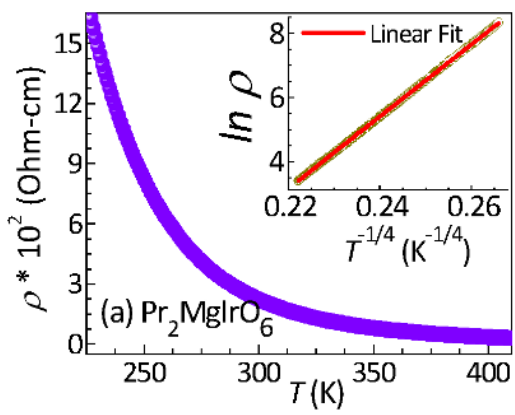
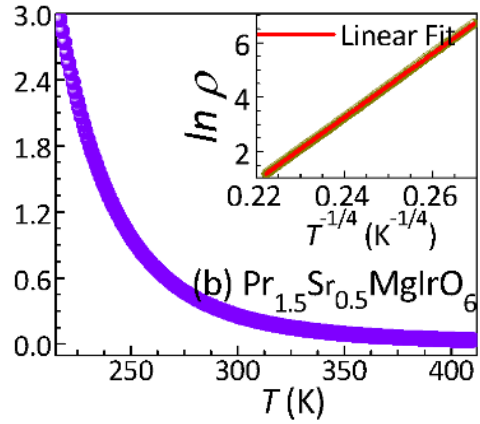
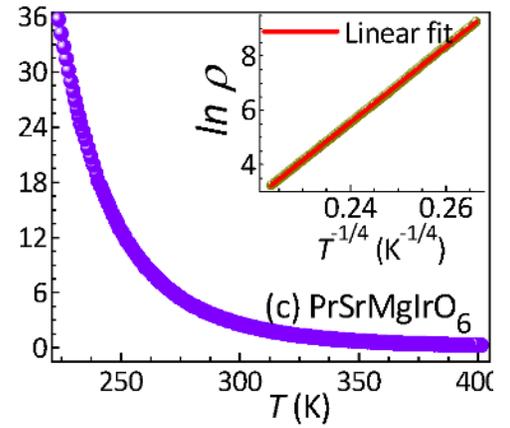
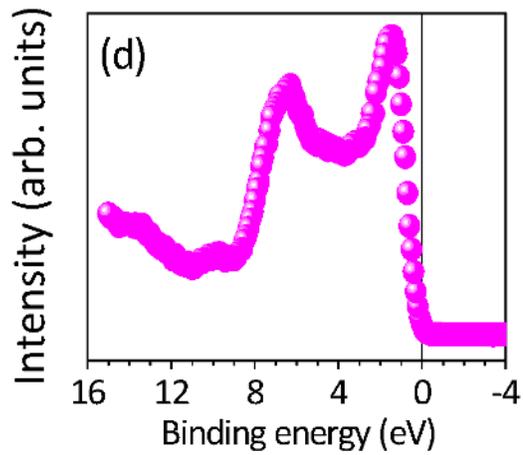
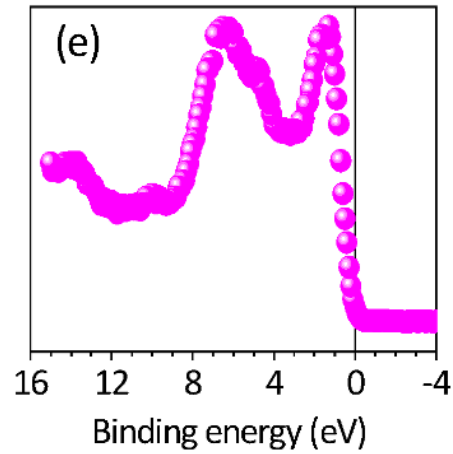
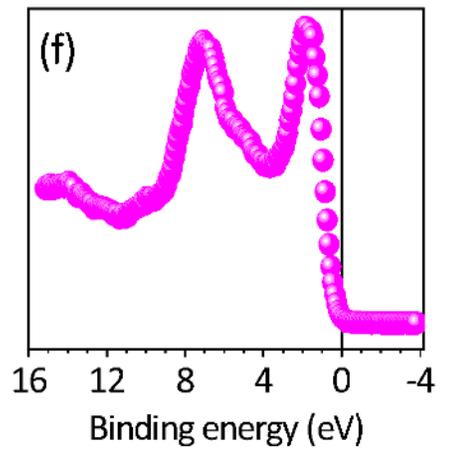
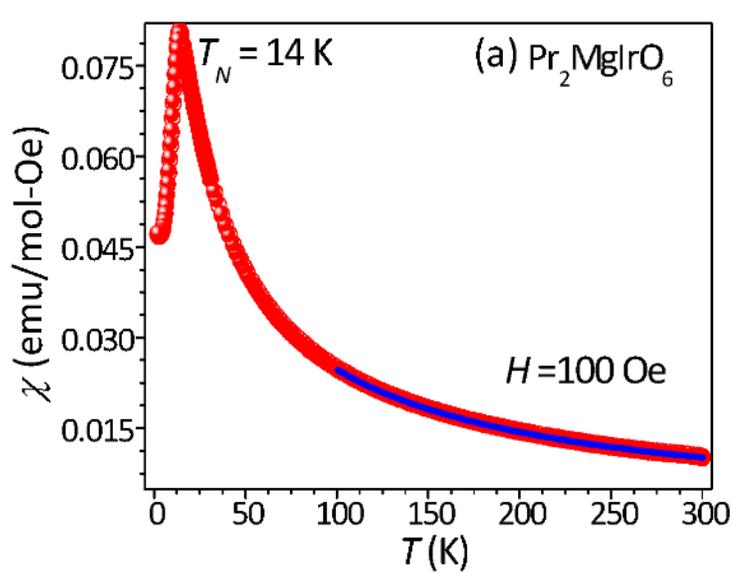
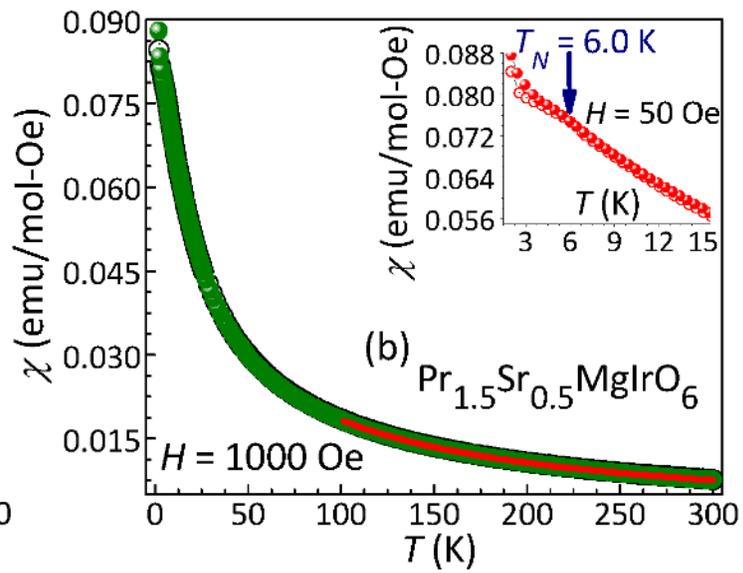
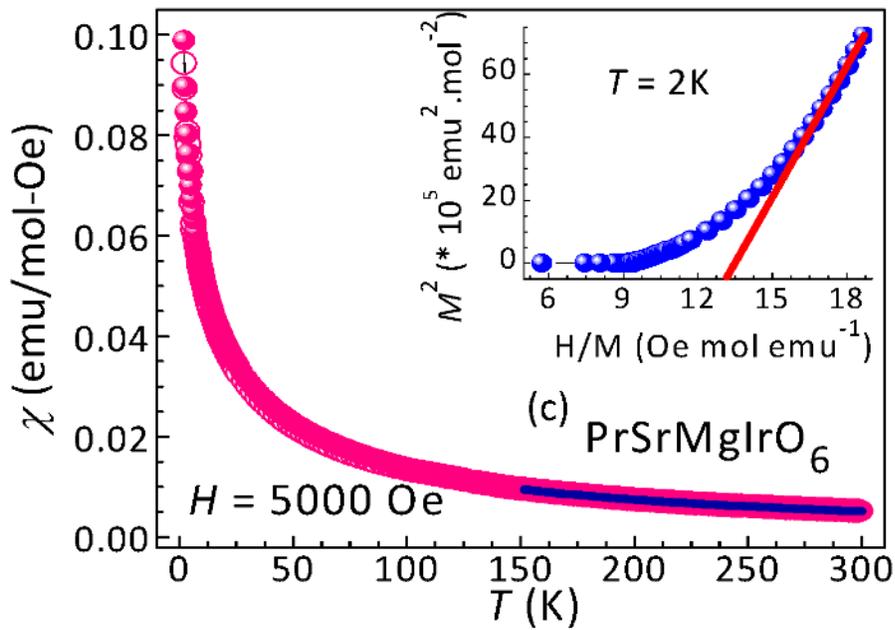
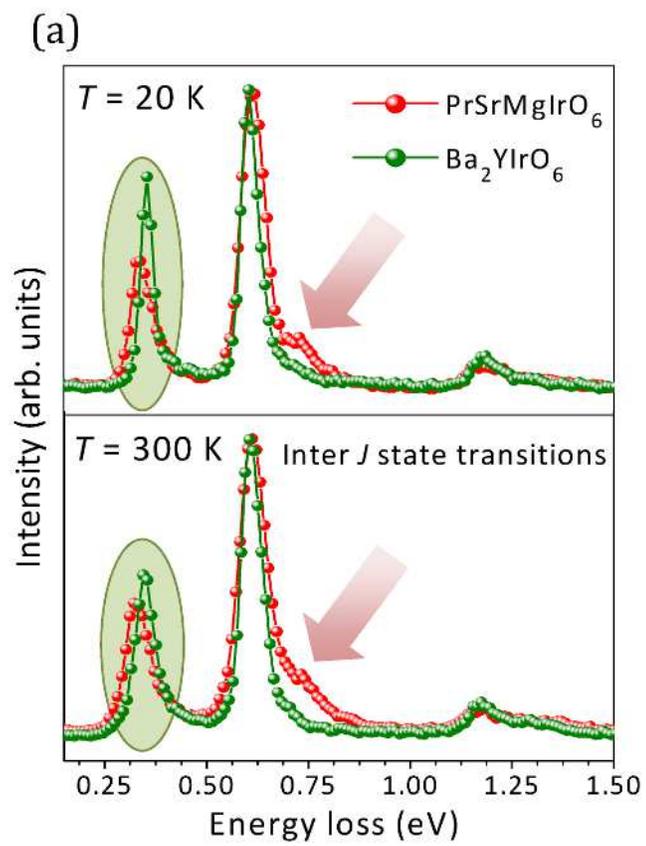
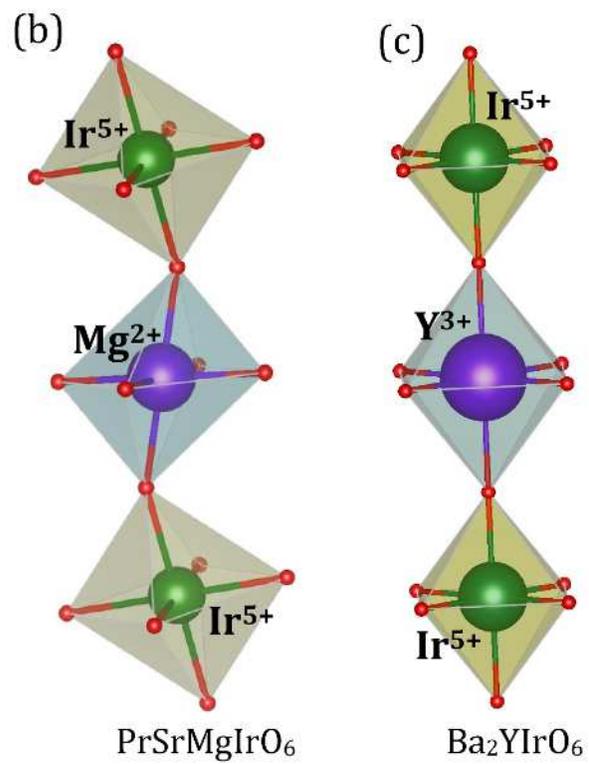
\end{document}